\documentclass[showpacs,nofootinbib]{revtex4} 
\usepackage{graphics}
\usepackage{color}

\usepackage{ulem}
\begin{document}
\title{Invariant color calculus and generalized Balitsky-Kovchegov hierarchy} 
\author{Alexey V. Popov} 
\email[]{avp@novgorod.net}
\affiliation{Novgorod State University, Velikiy Novgorod, Russia} 

\begin{abstract}
We derive generalization of the Balitsky-Kovchegov (BK) equation for a dipole, which consists of a parton and an antiparton of arbitrary charge.
At first, we develop one method of indexless transformation of color expressions. The method is based on an evaluation 
of the Casimir operator on a tensor product.
From the JIMWLK equation we derive the evolution equation for a single parton and prove gluon Reggeization in an arbitrary color channel.
We show that there is a color duplication of such Regge poles. 
Higher t-channel color exchange has its own Regge pole, which residue is proportional to the quadratic Casimir.
Taking a fundamental representation, we derive the usual BK equation and shed new light on the meaning of linear and nonlinear terms.
Finally, we discuss a linearized version of the generalized BK equation.
\end{abstract} 
\maketitle 
\section{Introduction} 
The Jalilian--Marian–-Iancu--McLerran-–Weigert-–Leonidov-–Kovner(JIMWLK) 
equation \cite{JIMWLK,CGC} is an important part of our understanding of high energy evolution of QCD scattering amplitudes.
It describes a scattering of a dilute projectile on a dense target. The main disadvantage is its functional form, which 
is necessary for studying of an arbitrary projectile. The functional equation is difficult to solve, even numerically.
The Balitsky-Kovchegov(BK) \cite{BK} hierarchy is a special case of the JIMWLK equation where the initial projectile is fixed and 
taken by a quark-antiquark pair. To perform practical calculations one can use the mean field approximation, which allows one to
reduce a full infinite hierarchy to a single closed equation, which can be solved both numerically \cite{BK_num} and analytically \cite{BK_an}.
From the mathematical viewpoint, derivation of the BK hierarchy is just a method of reduction of the JIMWLK Hamiltonian on some subspace of the functional space.
In this paper, we try to generalize such a method into a wider class of initial projectiles and consider a dipole that consists of a parton and an antiparton of arbitrary charge.
This step allows us to see the rich mathematical structure that arises in scattering amplitudes due to usage of the non-Abelian gauge group $SU(N)$ in QCD. 
A physical application of our method is an explicit demonstration of intensive color duplication of Regge poles in a scattering amplitude. 
Higher Pomerons, which are associated with higher representations of the gauge group, have been observed recently~\cite{Kovner_12}. 
They can arise in the analytical structure of the scattering amplitude when the initial projectile is more complex than an ordinary dipole.
In Ref. \cite{Kovner_12}, where a gluonic dipole was studied, the higher Pomerons were considered. 
The method, which was used in \cite{Kovner_12}, is bounded with specific properties of adjoint representation. 
In this paper, we propose a formalism that allows us to equally study a dipole which consists of a parton and an antiparton of arbitrary charge.
Mainly, we are interested in generalization of the BK hierarchy. The Pomerons can be analyzed in the weak--field limit of the BK equation.

Usually, in applications the Balitsky--Fadin--Kuraev–-Lipatov (BFKL) Pomeron \cite{BFKL} and its corrections are widely exploited. 
It is believed that the BFKL equation explains thr initial fast growing of gluon density at small $x$. 
However, poles with a higher color charge potentially may influence the calculation of experimental quantities.
From the theoretical viewpoint, a complete solution 
of evolution for an arbitrary projectile requires a formalism which can manipulate a contribution from any pole.

An additional interesting question is about theoretical aspects of the origin of the BFKL equation, especially the question about its domain of applicability. 
Let us describe shortly the current known approaches to the BFKL equation: 
\begin{itemize}
\item In a classical approach to the BFKL equation \cite{BFKL}, a Pomeron is considered as a bound state of two Reggeized gluons.
The disadvantage of this method is that it is not clear how to relate the BFKL equation with the full scattering amplitude of an arbitrary 
target and projectile. The bound state is constructed by hand, and there is no explicit algorithm of construction of the full amplitude.
A so-called Reggeon field theory should be developed.
\item We can use the dipole model where the limit $N\to\infty$ is assumed \cite{Dipoles}. In this limit,the BFKL equation arises without problems as the limit 
of a small amplitude of dipole scattering. The disadvantage of this method is that the limit $N\to\infty$ is strongly unnatural and
the rich algebraic structure associated with the realistic $SU(3)$ gauge group is lost. 
\item From the BK hierarchy we can derive the BFKL equation by applying the mean field approximation and the limit of weak dipole
scattering amplitude. The mean field approximation reduces the infinite hierarchy to a single nonlinear BK equation.
The disadvantages are that such an approximation has a limited domain of applicability and it is not clear how to generalize it to arbitrary initial 
projectiles. The advantage is only that target fields can be large without any restrictions. 
\item The weak--field approximation is an assumption that target field distribution is concentrated near zero field. 
The method allows us to study complicated projectiles and more complicated poles such as the odderon \cite{Odderon}.
We use this method in the current work in the process of linearization of a generalized BK equation. However, the general question here
is about the domain of applicability of the weak--field approximation. Since the target field is weak, the target must be dilute. Hence,
the dense-dilute picture is lost, and the JIMWLK equation is not applicable. 
In the dilute-dilute regime we must use another approach such as was proposed in \cite{Kovner_0501}.
However, in the current paper, we assume that the target field is small but is larger than the projectile field. 
So we use simultaneously both the JIMWLK equation and the weak field.
\end{itemize}

This paper is organized as follows. In Sec. \ref{sect_2} we develop the method of indexless transformation of color expressions, which is
widely used in the paper. The method is simple and is based on decomposition of the Casimir operator on a tensor product of two 
representations. It allows us to significantly simplify subsequent calculations. 
As an example of the power of the developed method, in Sec. \ref{sect_3} we apply it to make a fast calculation of the nontrivial color factor which is a convolution 
of eight structure constants. In Sec. \ref{sect_4} we take a single parton with arbitrary color charge, and 
from the JIMWLK equation we derive the single parton evolution equation and its restriction to the fundamental case. We observe that
this equation is a natural precursor of the common BK equation. 
In Sec. \ref{sect_5} from the single parton evolution equation we prove gluon Reggeization in an arbitrary color channel.
In Sec. \ref{sect_6} we derive a generalization of the BK equation by taking a dipole which consists of a parton and an antiparton of arbitrary charge.
Selecting fundamental representation, we easily reproduce the usual BK equation without any manipulation with color indexes. 
In Sec. \ref{sect_6a}, by using weak--field approximation, we show how to obtain the BFKL equation. 
Section \ref{sect_7} contains our conclusions.

\section{Casimir on a tensor product} \label{sect_2}
In many calculations in QCD we often need to transform a term like $T^a_R M T^a_R$, where $R$ is some color representation, 
$T^a_R$ is a generator, and $M$ is some matrix. Usually, such transformations are plagued by complicated index algebra. We develop the method 
of indexless decomposition that is based on the existence of the quadratic Casimir operator, which
for irreducible representations is proportional to the unit operator.

Consider a tensor product of two irreducible representations:  $A\otimes B$.
For the Lie group, generators in $A\otimes B$ have the form
\begin{equation} \label{eq_23}
T^a_{A\otimes B}=id \otimes T^a_B+ T^a_A  \otimes id
\end{equation}
The product $A\otimes B$ splits into a direct sum of irreducible representations:
\begin{equation}
A\otimes B=\bigoplus\limits_Q V_Q
\end{equation}
where $Q$ denotes representations and $V_Q$ corresponds to the invariant vector space. Generators also can be decomposed as
\begin{equation} \label{eq_6}
T^a_{A\otimes B}=\sum\limits_Q T^a_Q
\end{equation}
where $T^a_Q$ obey the usual reducibility conditions
$$ Im T^a_Q\subset V_Q $$
\begin{equation} \label{eq_7}
T^a_{Q_1}V_{Q_2}^{\phantom{a}}=0; \qquad \mbox{if $Q_1\neq Q_2$} 
\end{equation}
From definition (\ref{eq_23}) the
Casimir\footnote{In this paper we work only with the quadratic Casimir operator.} operator for $A\otimes B$ is
\begin{equation}
T^a_{A\otimes B}T^a_{A\otimes B}=C_A+C_B+2T^a_A \otimes T^a_B
\end{equation}
On the other side, from (\ref{eq_6}) and (\ref{eq_7}) we have
\begin{equation}
T^a_{A\otimes B}T^a_{A\otimes B}=\sum\limits_Q C_Q P_Q
\end{equation}
where $P_Q$ is a projection operator on subspace $V_Q$.
Index $Q$ runs over all irreducible representations which belong to the decomposition of the tensor product.
Projectors obey natural properties 
\begin{equation}\label{eq_20}
\begin{array}{l}
P_Q^2=P_Q \\
P_Q P_R=0; \quad \mbox{if $Q\neq R$} \\
Sp(P_Q)=D_Q
\end{array}
\end{equation}
where $D_Q$ is the dimension of representation $Q$.
Finally, we have 
\begin{equation} \label{eq_9}
2T^a_A \otimes T^a_B=\sum\limits_Q C_Q P_Q-C_A-C_B
\end{equation}
Now consider representation $R$ and any matrix $M$ that acts in $V_R$.
Under a gauge transformation matrix $M$ transforms as $M\rightarrow UMU^\dag$. Hence, it transforms as $R\otimes \bar R$, 
where $\bar R$ denotes complex conjugate representation (decomposition $R\otimes \bar R=1\oplus other$ gives 
the natural correspondence between forms on $V_R$ and vectors in $V_{\bar R}$). Small gauge transformations in $\bar R$ have the form
$\bar\psi \rightarrow \bar\psi (1-i\varepsilon^aT^a)$. We can conclude that $T^a_{\bar R}=-(T^a_R)^T$. Correspondingly, the term $T^a_R M T^a_R$
can be viewed as
\begin{equation} \label{eq_10}
 T^a_R M T^a_R=(-T^a_R \otimes T^a_{\bar R}) M
\end{equation}
The Casimir operator for complex conjugate representation is the same: $C_{\bar R}=C_R$. So with matrix notation we can write
\begin{equation} \label{eq_14}
 T^a_RMT^a_R=C_RM-\frac{1}{2} \sum\limits_Q C_Q M_Q
\end{equation} where $M_Q=P_Q M$. 

Equation (\ref{eq_14}) is useful for transformation of numerous terms in QCD calculations. Its sufficient advantages
are indexless and simple generalization for arbitrary representations. Widely used
in literature,\footnote{It is called the "Fierz identity"} simple version of (\ref{eq_14}) for fundamental representation is
\begin{equation} \label{eq_24}
T_{ij}^a T_{ks}^a=\frac{1}{2}\delta_{is}\delta_{jk}-\frac{1}{2N}\delta_{ij}\delta_{ks}
\end{equation}
We shall use (\ref{eq_14}) for derivation of a generalization of the BK equation for the dipole which consists of a parton and an antiparton of arbitrary charge.
In the next section we shall demonstrate the method by fast calculation of a cube diagram for gluons.

\section{Gluon cubic diagram} \label{sect_3}
\begin{figure}[h] 
\includegraphics{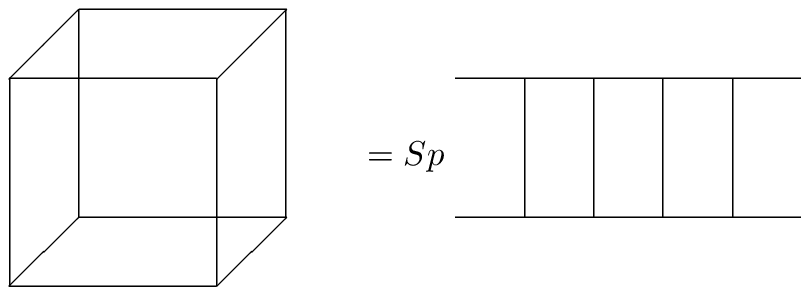} 
\caption{Cubic diagram. The vertex corresponds to structure constant $f^{abc}$, and the edge is an index convolution.}
\label{fig_1}
\end{figure}

There are works where universal tools for multigluon color factor calculations were studied \cite{ColorMangano,ColorDecomp}.
See also \cite{ColorFlow} for computer friendly color flow decomposition.
However, it would be suitable in partial situations to develop more special and simple method such as we have offered in the previous section.
As an example, we calculate the nontrivial color factor in the gluonic Feynman  
corresponding diagram from Fig. \ref{fig_1},
which is a convolution of eight structure constants 
\begin{equation} \label{eq_17}
I=f^{a_1a_2a_3}f^{a_2a_4a_5}f^{a_5a_9a_6}f^{a_3a_6a_7}f^{a_7a_8a_{12}}f^{a_8a_9a_{10}}f^{a_4a_{10}a_{11}}f^{a_1a_{11}a_{12}}
\end{equation}
It was calculated in Ref. \cite{GroupTheory} by a birdtrack method which is unavoidably lengthy in this case.
\noindent Of course, invariant (\ref{eq_17}) can be calculated using conventional properties of $f^{abc}$ and $d^{abc}$ symbols. 
Basic properties of these symbols can be found in the appendix to Refs. \cite{Kovner_12,NSZ}.
But this needs careful treatment of tensor expressions, especially the sign factor. Here we present the method of calculation in terms of 
$SU(N)$ invariants. 

It is clear from Fig. \ref{fig_1} that the value of (\ref{eq_17}) can be expressed as
\begin{equation} \label{eq_19}
I=Sp \left( T_8^a\otimes T_8^a\right)^4
\end{equation}
where we used the common fact that group structure constants are generators of the adjoint representation.
The following formula is a direct consequence of (\ref{eq_9})
\begin{equation} \label{eq_18}
T_R^a\otimes T_R^a=\frac{1}{2}\sum_Q C_Q P_Q - C_R=\sum_Q\left(\frac{1}{2}C_Q-C_R\right)P_Q
\end{equation}
By inserting (\ref{eq_18}) into (\ref{eq_19}) and using properties (\ref{eq_20}), we obtain
\begin{equation} \label{eq_21}
I = \sum_Q \left(\frac{1}{2}C_Q-C_8\right)^4D_Q
\end{equation}
where $D_Q$ is the dimension of the representation $Q$.
Now we need to know the properties of representations entered into the sum in (\ref{eq_21}). Decomposition of the tensor product
of two adjoint representations of $SU(N)$ group can be obtained using the Young diagrams. The Casimirs $C_R$ can be calculated in various ways \cite{Jeon_04}.
Here we only collect and write in Table \ref{tbl_1} the required values taken from \cite{Kovner_12}.

\begin{table}[h] \centering 
\begin{tabular}{|c|ccccccc|}
\hline
$R$ & 1 & $8_S$ & 27 & $R_7$ & $8_A$ & 10 & $\overline{10}$ \\
$D_R$ & 1 & $N^2-1$ & $\frac{N^2(N+3)(N-1)}{4}$ & $\frac{N^2(N-3)(N+1)}{4}$ & $N^2-1$ & $\frac{(N^2-4)(N^2-1)}{4}$ & $\frac{(N^2-4)(N^2-1)}{4}$ \\
$C_R$ & 0 & $N$ & $2(N+1)$ & $2(N-1)$ & $N$ & $2N$ & $2N$ \\ \hline
\end{tabular}
\caption{Properties of representations entered into (\ref{eq_21}).}
\label{tbl_1}
\end{table}

\noindent After some simple algebra we obtain the result
\begin{equation} \label{eq_22}
I=\frac{N^2}{8}(N^2+12)(N^2-1)
\end{equation}
There are two interesting consequences of (\ref{eq_22}). The first is the inapplicability of the large $N_c$ limit for the considered color factor
in the $SU(3)$ case due to factor $N^2+12$. The second consequence is the $SU(2)$-scaling 
violation\footnote{This clause is not about factor $N^2-1$, which is just the dimension of the adjoint representation.}.
This scaling is the observation that in many types of diagrams the color factor for the $SU(N)$ case is equal to the $SU(2)$-case factor multiplied
by some integer power of $(N_c/2)$.

The advantage of the presented calculations is the absence of any ugly tensor algebra. All relevant invariants for the $SU(N)$ group can be 
collected once into compact tables and repeatedly used later.

\section{Single parton evolution equation} \label{sect_4}
In many papers concerning the BK equation, authors usually start from the dipole scattering amplitude. However, we can also start
from the single quark scattering amplitude. In this paper, we consider the case when a projectile consists of only one parton in arbitrary 
color representation $R$. The high energy evolution of parton wave functions is governed by the
JIMWLK equation \cite{JIMWLK,CGC}. Here we use the notation taken from Ref. \cite{Kovner_08}. The JIMWLK equation is
 $$
  \frac{dS[\alpha] }{dY}=H[\alpha, \frac{\delta}{\delta \alpha}]S[\alpha]
 $$ $$
  H=\frac{g^2}{(2\pi)^3} \int\limits_{zxy}^{\phantom{x}}
  K_{zxy}
  \left[ -J^a_+(x)J^a_+(y)-J^a_-(x)J^a_-(y)+2V_{ba}(z)J^b_+(x)J^a_-(y) \right]  
 $$ 
 \begin{equation} \label{eq_1}
 K_{zxy}=\frac{(\vec z-\vec y)(\vec z-\vec x)}{(\vec z-\vec y)^2(\vec z-\vec x)^2} 
 \end{equation}
 $$
 J^a_\pm(x)=\frac{1}{ig}\frac{\delta}{\delta \alpha_a(x,\pm\infty)} 
 $$
 $$
 V(\vec x) = P e^{ig\int\limits_{-\infty}^{+\infty}\alpha_a(\vec x,x^+)T^a_{AD}dx^+} 
 $$
where $V_{ba}(z)$ is the gluon scattering amplitude in an external field $\alpha_a$, 
 $T^a_{AD}$ are generators of the gauge group in the adjoint representation, and 
$S[\alpha]$ is the projectile scattering amplitude as a functional of target fields. 
In order to obtain an observable scattering amplitude, we need to perform an average over target fields with a corresponding weight functional. 
However, for theoretical purposes it is useful to use a nonaveraged functional $S[\alpha]$ with fixed target fields.
Note that here we assume that the projectile is left-moving.

Now we would like to find the action of the JIMWLK Hamiltonian on a single parton S matrix. The latter is given by
\begin{equation}
S_R(\vec x)=P e^{ig\int \alpha_a(\vec x,x^+)T^a_R dx^+}
\end{equation}
where we assume that $S(\vec x)$ is a color matrix acting on the projectile color index. Functional derivatives can be easily 
evaluated:
$$J_+^a(x') S_R(x)=\delta(x'-x) T^a_R S_R(x)$$
\begin{equation}
J_-^a(x') S_R(x)=\delta(x'-x)  S_R(x) T^a_R
\end{equation}
where $T^a_R$ are the $SU(N)$ generators in representation $R$.
Convolution with $V_{ba}$ can be evaluated with the help of the following property which holds at any transverse point:
\begin{equation} \label{eq_28}
S^+T^a_RS=V_{ab}T_R^b
\end{equation}
After all evaluations we arrive at
\begin{equation} \label{eq_12}
\frac{dS_R(x)}{dY}=\frac{g^2}{(2\pi)^3}\int\frac{1}{(\vec z-\vec x)^2}\left[ 2S_R(z) T^a_R S_R^+(z) S_R(x) T^a_R-2C_RS_R(x)\right] d^2z
\end{equation}
Using decomposition (\ref{eq_14}), we have
\begin{equation} \label{eq_35}
\frac{dS_R(x)}{dY}=\frac{g^2}{(2\pi)^3}\int\frac{1}{(\vec z-\vec x)^2}
\sum_Q \frac{C_Q}{D_R}\left[ -S_R(z) \left(S_R^\dag(z)S_R(x) \right)_Q\right] d^2z
\end{equation}
In the fundamental representation we can simplify Eq. (\ref{eq_35}).
We know that $\bar 3\otimes 3=1\oplus 8$. The Casimirs are $C_1=0$ and $C_8=N$.
We have two projectors which obey $P_1+P_8=1$. A projector to an invariant state is $P_1M=Sp(M)/N$. So for any matrix $A$ we have
\begin{equation} \label{eq_36}
A_8=A-Sp(A)/N
\end{equation}
By substituting this into (\ref{eq_35}), we arrive at the first equation of "charged BK hierarchy"
\begin{equation} \label{eq_2}
\frac{dS(x)}{dY}=\frac{Ng^2}{(2\pi)^3}\int \frac{1}{(\vec z-\vec x)^2}
\left[ S(z)S(x,z)-S(x)\right]d^2z
\end{equation}
where
\begin{equation}
S(x,z)=\frac{1}{N}Sp(S^+(z)S(x))
\end{equation}
In the last equation we can easily recognize a dipole scattering amplitude. 
Equation (\ref{eq_2}) has a very natural and clear physical meaning. A quark at transverse position $\vec x$ emits a gluon into the
position $\vec z$.  The emitted gluon can be viewed as a quark-antiquark pair. The antiquark part of the gluon combines with the original
quark into a dipole. Then the first term in (\ref{eq_2}) can be viewed as multiple scattering of a 
dipole and quark component
of an emitted gluon. The second term corresponds to the virtual correction due to the requirement of the overall probability conservation.

It is not surprising that Eq. (\ref{eq_2}) is not closed.
This means that functional $S(x)$ does not form a complete space of solutions, and we must add 
corresponding equations for functional $S(z)S(x,z)$ and so on. So there is an infinite hierarchy of equations.
This hierarchy is just an attempt to reduce the JIMWLK Hamiltonian by some separable subspace of full functional space $S[\alpha]$.
If we take functional space $V$ such as $HV\in V$ then the JIMWLK evolution can be reduced by $V$.
The method of generation of hierarchy, like that started from (\ref{eq_2}), is the following.
For given starting space $V_0$ such as the space of $S(x)$ in (\ref{eq_2}), we generate space $V=\oplus_n H^nV_0$. 
It is clear that $HV\in V$, so the evolution can be reduced by $V$.

There are two approximate methods for closing an infinite hierarchy to a finite number of equations. The first method is the
mean field approximation where the target average leads to something like $\langle S(z)S(x,z)\rangle\to \langle S(z)\rangle\langle S(x,z)\rangle$.
The functional dependence on $\alpha_a(x)$ is removed, and we deal only with ordinary functions. 
The second method is the weak scattering approximation where we reduce full functional space by considering functionals only on small fields $\alpha_a(x)$.
Since the physical functional $S[\alpha]$ must obey $S[0]=1$, at small $\alpha_a(x)$ the functional $1-S[\alpha]$ is small, too.
We apply this method in the next section.

\section{Gluon reggeization} \label{sect_5}
Now we want to switch to the weak--field limit where it is assumed that the target fields are small. 
In this section, representation $R$ is assumed for symbols $S$ and $M$.
 In the weak scattering limit we define scattering amplitude $M(x)$ as
\begin{equation}
S(x)=1-M(x)
\end{equation}
In this limit $M(x)\ll1$, so we can 
keep in (\ref{eq_12}) only linear over $M$ terms  
and can use identity $M^\dag=-M$. By expanding (\ref{eq_12}), we obtain
\begin{equation} \label{eq_8}
\frac{dM(x)}{dY}=\frac{g^2}{(2\pi)^3}\int\frac{1}{(\vec z-\vec x)^2}\left[ 2C_R (M(z) - M(x)) - T^a_R (M(z) - M(x)) T^a_R\right] d^2z
\end{equation}
By substituting the expression (\ref{eq_10}) into (\ref{eq_8}) and making use of (\ref{eq_9}), we finally arrive at
\begin{equation} \label{eq_11}
\frac{dM_Q(x)}{dY}=\frac{C_Q g^2}{(2\pi)^3}\int\frac{1}{(\vec z-\vec x)^2}\left[ M_Q(z) - M_Q(x)\right] d^2z
\end{equation}
for each irreducible representation $Q$ which contributes to a tensor product $R\otimes \bar R$. The matrixes $M_Q$ are
$M_Q=P_Q M$ as it was defined in (\ref{eq_14}).

Equation (\ref{eq_11}) provides the closed expression for the evolution of the scattering amplitude of the projectile
in the given representation $R$ in the channel with the given color exchange $Q$. To find a complete solution, we need to decompose the initial
conditions of $M$ to a sum of irreducible representations and solve for each component Eq. (\ref{eq_11}). 
Note that the singlet representation has zero Casimir, and it gives a constant solution of (\ref{eq_11}).

Equation (\ref{eq_11}) can be easily solved in the momentum space. By converting (\ref{eq_11}) to the momentum space, we obtain
$$\frac{dM_Q(k)}{dY}=\omega_Q(k) M_Q(k)$$
\begin{equation} \label{eq_4}
\omega_Q(k)=\frac{C_Qg^2}{(2\pi)^3}\int\frac{e^{-i\vec k\vec z}-1}{z^2}d^2z
\end{equation}
The solution can be easily obtained as
\begin{equation}
M_Q(k,Y)=M_Q^{(0)}(k)e^{\omega_Q(k)Y}
\end{equation}
where $M_Q^{(0)}=P_Q M^{(0)}$ are initial conditions of evolution.
This solution has Regge form $M\sim s^{\alpha(t)}$ as it should. 
If $Q$ equals the adjoint representation, then the known expression of a gluon pole trajectory has the form \cite{Balitsky_01}
\begin{equation} \label{eq_5}
\alpha(k)=-\frac{\alpha_sN}{4\pi^2} \int \frac{k^2}{ p^2(\vec k-\vec p)^2}d^2p
\end{equation}
Integrals in (\ref{eq_4}) and (\ref{eq_5}) can be evaluated and they give equivalent answers
\begin{equation} \label{eq_25}
\omega_Q(k)=-\frac{C_Q\alpha_s}{2\pi}\ln \frac{k^2}{\mu^2}
\end{equation}
where $\mu$ is an infrared regulator.

There is one important subtlety here. Though in the weak--field limit we require $M\ll 1$,
individual components of $M$ may have distinct order in comparison with each other.
Indeed, let us recall that $M=1-\exp(i\alpha_a T^a)$. Since there are two obvious identities
$(T^a)_8=T^a$ and $(T^a)_{\neq 8}=0$ (index 8 denotes here the adjoint representation that is constructed from 
generators), we have
\begin{equation} \label{eq_37}
\begin{array}{l}
M_{\neq 8}=O(M_8^2) \\
M\simeq M_8
\end{array}
\end{equation}
This means that the main contribution in (\ref{eq_11}) comes from the adjoint representation with $C_8=N$. 
The other $M_Q$ has order at~least $(M_8)^2$, which is negligible in the weak--field limit.
However, we can study Eq. (\ref{eq_11}) beyond the weak--field limit, too.
It may have sense as a term of formal power expansion of the full evolution equation (\ref{eq_35}). 
This expansion may be useful for construction of something like the Reggeon diagram technique. 
Also, higher representations $Q$ may be relevant when the target has specific field correlators which 
allow one to set $\langle M_8 M_8\rangle \simeq 0$ and similar for higher powers.

The evolution equation (\ref{eq_4}) corresponds to a single moving pole in a complex angular momentum plane. 
The pole trajectory is given by solution (\ref{eq_25}). So we can see the so-called reggeization phenomenon. However, we
have found more than one pole -- one pole for each irreducible representation $Q$. This is a really remarkable result since 
there is a transfer of gauge group algebraic structures to the analytic structure of the scattering amplitude.
Conversely, from the analytic structure of a scattering amplitude we can enumerate irreducible representations.
A similar situation is observed in the Pomeron trajectory \cite{Kovner_12}. Since the Pomeron can be viewed as a bound state of two
Reggeized gluons, we naturally conclude that there is an intensive color duplication of poles.
Poles with higher $Q$ can be relevant where there are many native partons (quarks or gluons) in a projectile.
Several native partons can form a higher representation via the tensor product of their color spaces.
The significant feature of the BFKL Pomeron case is that there are many poles even without color duplication.
This happens due to strong degeneration of the eigenvalues of the spectrum of the BFKL operator.

\section{Generalized BK equation} \label{sect_6}
 Now we consider the case where the projectile is a color dipole built from two partons in representations $R$ and $\bar R$. We want to 
 study evolutions of the following scattering functional: 
\begin{equation} \label{eq_27}
 S=\frac{1}{D_R} Sp\left(U^\dag(y) U(x)\right)
\end{equation}
where $U$ is the Wilson line of representation $R$. 
Since the JIMWLK operator $H$ in (\ref{eq_1}) is a differential operator of second order, in calculation of $HS$
we can apply the Leibnitz rule, which gives four terms which are equal to each other but with different spatial kernels.
We show the term where all $J^a$ act on the $U(x)$ in (\ref{eq_27}). It is 
\begin{equation} 
\frac{1}{D_R}Sp(U^\dag_yHU_x)=\int_z
\frac{g^2K_{zxx}}{(2\pi)^3D_R}Sp\left( 
- U_y^\dag T^a_R T^a_R U_x - U_y^\dag U_x T^a_R T^a_R + 2V_{ba} U_y^\dag T^b_R  U_x T^a_R
\right)
\end{equation}
The calculation of the other three terms is very similar. By using $T^a_RT^a_R=C_R$ and the identity (\ref{eq_28}), which is read as
\begin{equation}
  V_{ba}(z)T^b=U(z)T^aU^\dag(z)
\end{equation}
and adding three remaining terms, we obtain
 \begin{equation} \label{eq_13}
 \frac{dS}{dY}=HS=\frac{g^2}{(2\pi)^3}\int\limits_{z} M_{zxy} \frac{1}{D_R} Sp\left[2U^\dag_yU_zT_R^aU^\dag_zU_xT_R^a-2C_RU^\dag_yU_x\right]
 \end{equation}
 where $M_{zxy}$ is the well-known dipole kernel
 \begin{equation}
 M_{zxy}=K_{zxx}+K_{zyy}-K_{zxy}-K_{zyx}=\frac{(x-y)^2}{(z-x)^2(z-y)^2}
 \end{equation}
 The key point of our method is the usage of decomposition (\ref{eq_14}), which in the current context has the form
\begin{equation} \label{eq_26}
 T_R^aU^\dag_zU_xT_R^a=C_RU^\dag_zU_x-\sum_Q\frac{1}{2}C_Q\left(U^\dag_zU_x\right)_Q
\end{equation}
As usual, $Q$ runs over irreducible representations which contribute to the tensor product $R\otimes \bar R$.
By substituting (\ref{eq_26}) into (\ref{eq_13}), we finally arrive at 
 \begin{equation} \label{eq_15}
 \frac{dS}{dY}=\frac{g^2}{(2\pi)^3}\int\limits_{z} M_{zxy} \sum\limits_Q \frac{C_Q}{D_R} Sp\left[-U^\dag_yU_z\left(U^\dag_zU_x\right)_Q\right]
 \end{equation}
 When $R$ equals the fundamental representation, we can easily obtain the usual BK equation. Using Eq. (\ref{eq_36}),
 we obtain a simplified version of (\ref{eq_15}):  
 \begin{equation} \label{eq_16}
 \frac{dS(y,x)}{dY}=\frac{g^2N}{(2\pi)^3}\int\limits_{z} M_{zxy} [S(y,z)S(z,x)-S(y,x)]
 \end{equation}
Equation (\ref{eq_15}) is the first equation of a complicated nonlinear hierarchy which has an additional complexity level 
in comparison with the usual BK hierarchy. This happens due to the presence of the sum over various $Q$ projections. 
When we further calculate $HS_{yz}S_{zx}$ we obtain new functionals which are various combinations of the Wilson lines,
and when we perform a corresponding operation on right-hand side of (\ref{eq_15}) we obtain such functionals
with two various $Q$ projections.
It should be stressed that linear and nonlinear terms in (\ref{eq_16}) are single whole which is just a projector on the adjoint component 
of the tensor product. This presents a contrast to the common description where the linear term is explicated by so-called virtual corrections.

\section{BFKL equation} \label{sect_6a}
At first, let us show shortly how to obtain the BFKL equation from (\ref{eq_16}) by using the mean field approximation as mentioned in the introduction.
We should not forget about target averaging in (\ref{eq_16}). 
The approximation gives $\langle S(y,z)S(z,x)\rangle=\langle S(y,z)\rangle\langle S(z,x)\rangle$. Next, 
we define $\langle S \rangle=1-N$. By taking the limit $N\ll 1$, we obtain the usual BFKL equation.  
It should be stressed that claim $N\ll 1$ does not assume a small target field. 
Unfortunately, it is not clear how to generalize the considered method to a case of more complicated projectiles.
 
In order to obtain the generalized BFKL equation, we consider the weak scattering limit of (\ref{eq_15}).
Let $U=1-M$. Since $U\in SU(N)$, we have
\begin{equation} \label{eq_29} 
M^\dag+M=MM^\dag
\end{equation}
Since $M_1=Sp(M)/D_R$ and $C_1=0$, we have the following useful relation for any $M$ and $Q$:
\begin{equation}  \label{eq_30} 
C_Q Sp(M_Q)/D_R=C_Q (M_Q)_1=0 
\end{equation} 
and there is a similar fact about the unit matrix 
\begin{equation} \label{eq_32}
C_Q 1_Q=0
\end{equation}
By expanding up to second order the matrixes in the right-hand side of (\ref{eq_15})  over $M$, we have 
\begin{equation}
\begin{array}{rl}
U^\dag_yU_z\left(U^\dag_zU_x\right)_Q=& 1_Q-M^\dag_y 1_Q-M_z 1_Q-(M^\dag_z)_Q-(M_x)_Q+ \\
&+M^\dag_yM_z 1_Q+M^\dag_y(M^\dag_z)_Q+M^\dag_y(M_x)_Q+M_z(M_z^\dag)_Q+M_z(M_x)_Q+(M_z^\dag M_x)_Q
\end{array} \end{equation}
Next, using (\ref{eq_29}),(\ref{eq_30}), and (\ref{eq_32}) we obtain
\begin{equation} \label{eq_31}
\frac{dN(y,x)}{dY}=
\frac{g^2}{(2\pi)^3}\int\limits_{z} M_{zxy} \sum\limits_Q C_Q
\left[ N_Q(y,z)+N_Q(z,x)-N_Q(y,x)-N_Q(z,z)\right]
\end{equation}
where we have defined
\begin{equation}
N_Q(y,x)=1-S_Q(y,x)=1-\frac{1}{D_R}Sp(U^\dag(y)U_Q(x))=\frac{1}{D_R}Sp(M^\dag (y)1_Q+M_Q(x)-M^\dag (y) M_Q(x))
\end{equation}
It can be easily checked that the linear over $M$ terms cancel each other in (\ref{eq_31}).

It is instructive to obtain the original  BFKL equation from (\ref{eq_31}).
Note that before this point we used only the assumption $M \ll 1$. 
However, as was shown in (\ref{eq_37}) of Sec. \ref{sect_5}, in the weak--field limit
only the $M_8$ component is relevant. The other $M_Q$ has order at least $O(M_8^2)$.
Hence, the sum over $Q$ in (\ref{eq_31}) is reduced to one term with $Q=8$. In additional, 
properties (\ref{eq_37}) allow one to set $N_8(y,x)=N(y,x)$. So we obtain the original BFKL equation
\begin{equation}
\frac{dN(y,x)}{dY}=
\frac{g^2N_c}{(2\pi)^3}\int\limits_{z} M_{zxy} \left[N(y,z)+N(z,x)-N(y,x)\right]
\end{equation}
where we used $N(z,z)=0$, since $U_z U_z^\dag=1$ exactly.

Beyond the weak--field limit, from a power expansion viewpoint, 
Eq. (\ref{eq_31}) inevitably contains the term $N_Q(z,z)$. 
Note that a similar term is already known. In Ref. \cite{Odderon} it arises in the evolution equation for the 2-point Green's function in the weak--field regime.
The functional $N_Q(z,z)$ depends on the field $\alpha_a(z)$ and can be viewed as a function of one variable on a group manifold. 
The mapping $\alpha_a \to SU(N)$ is $\exp(i\alpha_aT^a)$. This function on a group manifold is intrinsic, like characters, in the sense 
that it can be constructed directly from the definition of a group.
The system of equations (\ref{eq_31}) is not closed because there are many unknown variables $N_Q$ in the right-hand side. 
This means that there are exchanges of many different poles in the dipole scattering amplitude. 
To find the color diagonal version of (\ref{eq_31}), we must consider the action
of the JIMWLK operator $H$ on a two parton matrix element $U^\dag_y\otimes U_x$ with arbitrary color indexes and perform 
a diagonalization of color structure. In particular, we can show that the system of equations which is obtained from evaluation of
$HSp(U^\dag_y(U_x)_Q)$ in the weak scattering approximation is linear and closed. This means that the evolution equation for $N_Q(y,x)$ 
contains only other $N_{Q'}(y,x)$. Unfortunately, this system is intricate in comparison with the original BFKL equation.
See Ref. \cite{Kovner_12} for the eigenvalue problem in the adjoint case and Ref. \cite{NSZ} for a detailed study of multigluon states.

\section{Conclusion} \label{sect_7}
In this paper, we have studied one relation of gauge group algebraic structures to the high energy QCD evolution.
The classification of classical Lie groups and its irreducible representations is the classical mathematical result which is widely
used in modern theoretical physics. We are convinced that a full solution of high energy QCD evolution must be naturally related
to the rich mathematical structures on the gauge group. In fact, many known investigations ignore such structures.
We have seen that the form (\ref{eq_16}) of the usual BK equation is not natural and the more natural form is Eq. (\ref{eq_15}). 
Moreover, it is clear that the initial choice of a color dipole is not natural either. More simple and transparent equations 
emerge when we start evolution from a single parton. Geometrically, such choices of initial states are equivalent to selecting
finite polynoms of matrix elements of irreducible representations as an initial condition of functional $S[\alpha]$. It is clear
that during evolution the power of such polynoms is growing. So it is difficult to find appropriate analytic solutions for all $Y$.
Currently, our road map is to work on a functional level \cite{avp_extremal}. 
Powerful mathematical techniques can be used if we view $S[\alpha]$ as functions on the group manifold.

One can think that higher representations are irrelevant to the real world, where we deal only with quarks and gluons.
However, if a projectile has many partons and we use coarse transverse resolution, then 
few partons via tensor product can form a new effective parton, which belongs to the representation of a higher color charge.
This fact was widely used in Ref. \cite{Jeon_04}, where it was shown that in the dense case the charges with higher Casimir become dominant.

\section*{Acknowledgments}
We thank N.V. Prikhod'ko for feedback and useful remarks.

\end{document}